\documentstyle[preprint,aps,prb]{revtex}
\begin{document}
\draft
\preprint{}

\title{Non--Nominal Value of the Dynamical Effective Charge in
Alkaline--Earth Oxides}
\author{M. Posternak\cite{irrma} and A. Baldereschi\cite{irrma}}
\address{Institut de Physique Appliqu\'ee, Ecole
Polytechnique F\'ed\'erale, PHB Ecublens, CH--1015 Lausanne,
Switzerland}
\author{H. Krakauer}
\address{Department of Physics, College of William and Mary,
Williamsburg, Virginia 23187--8795}
\author{R. Resta}
\address{INFM--Dipartimento di Fisica Teorica, Universit\`a  di
Trieste, Strada Costiera 11, I--34014 Trieste, Italy}
\date{\today}
\maketitle
\begin{abstract}
We calculate {\it ab--initio} the electronic states and the Born
dynamical charge $Z^*$ of the alkaline--earth oxides in the
local--density approximation. We investigate the trend of increasing
$Z^*$ values through the series, using band--by--band decompositions
and computational experiments performed on fake materials with
artificially--modified covalence. The deviations of $Z^*$ from the
nominal value 2 are due to the increasing interaction between O $2p$
orbitals and unoccupied cation $d$ states. We also explain the
variations, along the series, of the individual contributions to $Z^*$
arising from the occupied band manifolds.
\end{abstract}


\narrowtext

The dynamical charges (also called Born effective charges or transverse
charges) measure the macroscopic polarization induced by relative
sublattice displacements in a polar insulator. When the ionic material
is partly covalent, the dynamical charges present nontrivial features:
the displacement of a given ion induces a substantially nonrigid
displacement of the associated electronic charge. This case includes
oxides, which have received recently attention from different groups.
Apparently similar oxides have strikingly different behavior: some
(like the ferroelectric perovskites) have giant dynamical
charges,\cite{perov,Ghosez95} while others (like ZnO) display dynamical
charges close to  the nominal static ones.\cite{rap88} The present work
addresses the series of the alkaline--earth oxides MgO, CaO, SrO, and
BaO. They have the rocksalt structure, and therefore the anion and
cation Born tensors are diagonal, with opposite values of the charges.
Their dynamical charges $Z^*$ show experimentally a well defined trend
in their deviation from the nominal value of the rigid--ion picture,
{\it i.e.} $+$2 (cation) and $-$2 (anion) in units of the elementary
charge, which will be used throughout. Lattice--dynamical measurements
of the LO-TO splitting at the Brillouin-zone center allow an accurate
determination of $|Z^*|^2/\varepsilon_\infty$, where
$\varepsilon_\infty$ is the electronic (``static--high--frequency'')
dielectric constant and is the major source of uncertainty when
comparing theory with experiment.  For MgO, the observed $Z^*$ is
nearly nominal; in the heavier oxides $Z^*$ increases monotonically
along the series, up to the value 2.7--3.0 in BaO. In the present
theoretical study we investigate such a trend, focussing in particular
on the individual contributions due---in each material---to different
groups of occupied bands. As an additonal tool, we also perform
computational experiments on ``fake'' materials, where the strength of
covalence has been artificially reduced.

The macroscopic polarization induced by a given sublattice displacement
in polar dielectrics is related to the integrated macroscopic current
traversing the sample. From a theoretical standpoint, reasoning in
terms of {\it currents}---as opposite to {\it charges}---provides bulk
quantities which are well defined in the thermodynamic limit and solves
many of the problems which have plagued the polarization concept until
recently.\cite{modern,rap_a12} The Born dynamical--charge tensors
measure the relationship between sublattice displacements and the
corresponding macroscopic electric fields. The Born charge of a given
ion is also defined as the second derivative, with respect to the field
and to the displacement, of a suitable thermodynamic potential (see
e.g. Ref.~\onlinecite{rap_a12}, Sec. VIII). As such, $Z^*$ measures the
linear polarization induced in zero field by a unit sublattice
displacement, or equivalently the linear force induced by a unit field
upon a given ion, at zero displacement. The modern theory of the
macroscopic polarization---implemented in the present study with
first--principles methods---allows one to calculate Born charges as
Berry phases.\cite{modern,rap_a12}

All our calculations are semirelativistic and use the self--consistent
full--potential linearized--augmented--plane--wave (FLAPW)
method.\cite{lapwref} The local--density approximation (LDA) to the
density functional theory with the Hedin--Lundqvist exchange and
correlation potential has been used. Through this study, we have used
the experimental lattice constants of the materials, whose values in
a.u. are 7.97  (MgO), 9.09 (CaO), 9.75 (SrO), and 10.46 (BaO). In the
following, we will indicate generically as MO any one of these four
oxides.

The accurate evaluation of the MO wavefunctions, as required in the
calculation of Berry phases,\cite{modern,rap_a12} is a challenge for
all band structure methods, including pseudopotentials, because of the
large energy range that must be spanned by the basis functions.  All
occupied states ranging from cation M n$p$ (Mg: n=2; Ca: n=3; Sr: n=4;
Ba: n=5) and oxygen $2s$ up to oxygen $2p$ have {\it a priori} to be
retained, leading, {\it e.g.}, to an energy range of 39 eV for MgO.
Specific to the LAPW method is the fact that O $2p$ tails extending
inside the cation muffin--tin spheres have to be described very
accurately. This is a difficult task in the standard LAPW method
because linearization is performed around M n$s$ and M n$p$ energies
inside cation spheres, {\it i.e.} much below the relevant O $2p$
energies. Using two energy windows with different linearization
energies and diagonalizations in each window doesn't solve entirely
the problem. Indeed, explicit interwindow orthogonalization is usually
not implemented in the computer programs, and non--orthogonality
effects corrupt substantially the overlap matrices required for the
Berry phase calculation. One way to circumvent these difficulties is to
expand the usual LAPW basis set with local orbitals.\cite{Singh91}
Mutual state orthogonality is then enforced, and variational freedom is
increased. An alternative way, consists simply of shrinking the cation
sphere until the largest part of the O $2p$ orbitals is in the O
spheres and in the interstitial region. A standard one--window LAPW
calculation is then performed, with an increased cutoff in reciprocal
space corresponding to the shrinking of the cation sphere. We have used
the latter approach in our study, keeping the oxygen sphere radius
fixed to the value of 1.8 a.u. The above criteria for the cation
spheres lead to the values 1.1, 1.6, 1.9, and 2.3 a.u. for the Mg, Ca,
Sr, and Ba sphere radii, respectively. We have checked that results are
in good agreement with those obtained from the LAPW/local--orbital
method and the larger values of the cation radii (1.5 a.u. in the case
of MgO).

We show in Fig.~\ref{fig1}a the electronic energy bands of MgO, and in
Fig.~\ref{fig2} those of the other materials. Several features of the
band structures have some relevance to the analysis of the dynamical
charges, and therefore we pause to scrutinize them in some detail.

Starting with the O $2s$ band, we notice that its center is about 15 eV
below the edge of the O $2p$ valence bands in all four materials. This
finding agrees with experiment though the magnitude is less than the
measured one,\cite{refxps} 18--21 eV, as a consequence of the well
known deficiencies of LDA. By contrast, the position of the M n$p$
bands varies substantially. For MgO these bands are about 39 eV (to be
compared with the experimental value\cite{refxps} $\sim46$ eV) below
the valence--band edge and are not visible in Fig.~\ref{fig1}a. In CaO
they appear at the bottom of the figure, in SrO they overlap with the O
$2s$ band, and in BaO they are a few eV above it. This trend agrees
with the experimental findings, despite smaller values of the energy
separation from the valence edge. In all cases except MgO, while there
is a non--negligible hybridization between O $2s$ and M n$p$ states,
reaching a maximum in SrO, the band complex O $2s$ + M n$p$ scarcely
interact with the remaining occupied states. We thus anticipate that
the contribution to $Z^*$ from this complex will be close to the
nominal value, while the individual O $2s$ and M n$p$ contributions
will deviate substantially.

There are also conspicuous conduction--band differences between MgO and
the other oxides, which can be understood easily in terms of M n$d$
states. For Mg (n=2), these states are absent and the gap is direct,
whereas for n$>$2 they contribute significantly to the lowest
conduction bands, leading to an indirect gap at X. The lack of $d$
states at low energy in MgO is also responsible for a lattice constant
significantly smaller than in the other compounds. In order to separate
a pure volume effect, we have performed an additional self--consistent
band--structure calculation for MgO under a huge negative pressure,
such that the lattice constant is increased up to that of CaO: these
bands are shown in Fig.~\ref{fig1}b. The energy gap decreases, the
lowest conduction bands remain predominantly Mg $3s$, O $3s$, and O
$3p$ and display a free--electron like behavior, while the highest O
$2p$ valence bands flatten as a consequence of reduced interatomic
interactions.

The three heaviest oxides have qualitively similar band structures, in
particular the highest valence bands, and a gap which decreases along
the series. The highest valence bands are nominally O $2p$, but
inspection of the wavefunctions shows a partial $d$ character (about
one order of magnitude larger than in MgO), hence a sizeable bonding
with the M n$d$ orbitals which determine the conduction band minimum at
X.

In the following, we will study the dynamical charge $Z^*_{\rm O}$
dragged by the displacement of an oxygen atom only. We have however
explicitely verified that in our Berry phase calculation the acoustic
sum rule $Z^*_{\rm M} + Z^*_{\rm O} = 0$ is fulfilled with great
accuracy through the whole series. The results are reported in
Table~\ref{table1} and compared to experimental\cite{landolt82} data,
whose range is due to uncertainty in the dielectric constant
$\varepsilon_\infty$. For the whole series the agreement of the
theoretical results with the experimental ones is very good. Our
results also agree with those of an ab--initio calculation for MgO,
CaO, and SrO,\cite{Schutt94} based on pseudopotentials and
linear--response theory: notice that in this paper the Szigeti charges
$Z^*_S = 3Z^*/\varepsilon_\infty$ are reported. For the heaviest
compound, BaO, we have checked that inclusion of the spin--orbit
coupling doesn't modify significantly the calculated value of 
$Z^*_{\rm O}$.

In our approach we can also calculate unambiguously the separate
contributions to $Z^*_{\rm O}$ from different electronic states,
whenever the corresponding bands are well separated in energy. Such a
decomposition, performed previously in other
materials,\cite{Ghosez95,rap88} is of paramount importance in
identifying the physical mechanisms responsible for the deviations of
$Z^*$ from the nominal ionic values.

The oxygen $1s^2$ state can be safely considered as rigid, so the ionic
core carries a charge of $+$6. In a completely ionic picture the
electronic states $2s$ and $2p$ are fully occupied with eight
electrons, thus providing the total nominal charge of $-2$. This
oversimplified picture must be replaced, for real materials, by the
values reported in Table~\ref{table1}. The polarizable electronic
states that can in principle contribute non trivial values, are M n$s$,
M n$p$, O $2s$, and O $2p$. For all materials, except SrO, the
corresponding bands do not overlap with each other and their
contribution can be computed separately. It is apparent in
Table~\ref{table1} that the M n$s$ states, which are very deep in
energy and localized in space, can be considered as rigid, {\it i.e.}
the physics of the dynamical charge is in the behavior of the other
three groups of bands.

Let us start with MgO, which behaves differently from the rest of the
series. The O $2s$ states are rather far from any other electronic
state, but nonetheless their share of dynamical charge is $-2.21$,
larger than the nominal value. A similar peculiar feature of the O $2s$
states was found for other materials\cite{Ghosez95,rap88} but no
satisfactory explanation of this finding has been proposed so far:
therefore we scrutinize the issue in more depth here. First we notice
that the Mg $2p$ states have a negligible contribution, as expected
from their energy location. Secondly we notice that the O $2p$ states
contribute less than the nominal value, in such a way that the effects
of the only relevant bands (oxygen $2s$ and $2p$) add up to a total
dynamical charge close to the nominal value. The reason for this is an
appreciable interaction between O $2s$ and O $2p$ states, which we have
identified by inspection of the relevant Bloch functions, and which is
basically due to the short nearest--neighbor distance in MgO. We have
further checked this by performing a separate calculation for a fake
MgO, where we have added a repulsive potential, confined to the oxygen
spheres, and acting only upon the radial $s$ wavefunction. With a
repulsive strength of 0.2 Ry, the $2s$ and $2p$ energy bands become
closer in energy, and their partial contributions to $Z^*_{\rm O}$
become $-$2.25 ($2s$) and $-$5.82 ($2p$). The total value of $Z^*_{\rm
O}$ is essentially identical for the fake and for the real material,
which demonstrates our point. MgO is therefore a material where the
occupied O $2s$ and O $2p$ orbitals interact amongst themselves, but
scarcely with empty conduction states: when oxygen is displaced, the
strength of the O $2s$ $-$ O $2p$ interaction is affected, but the
subspace of the corresponding occupied bands remains about the same.

The other materials are---as expected---qualitatively different from
MgO, and more similar amongst themselves. In agreement with the trends
detected in the band structure analysis, the interaction of O $2s$ and
M n$p$ states is significant, and results in a strongly non--nominal
value of the partial charges associated with the corresponding bands.
Looking first at the partial contribution due to O $2s$ states (in CaO
and BaO) we notice a much larger deviation from the nominal value than
in MgO. We understand this as the interaction with M n$p$ bands; in
addition, there is an interaction with O $2p$ states, similar to the
one occurring in MgO, although much weaker. However, in the three
heavier materials, the O $2s$ and M n$p$ contributions add up to an
almost nominal value: $-$2.09 in CaO, $-$2.06 in SrO, and $-$1.98 in
BaO. Therefore, the deviation of $Z^*_{\rm O}$ from the nominal value
is due to the partial contribution of the highest occupied bands, which
are O $2p$ with substantial bonding to cation n$d$ states. The
phenomenon here is then qualitatively similar to---although
quantitatively less spectacular than---the one we have discovered
previously\cite{perov,Ghosez95} in ferroelectric KNbO$_3$.

In order to substantiate this claim, we have performed a series of
calculations on fake materials, where the bonding of O $2p$ states with
cation n$d$ states is artificially suppressed by adding a repulsive
potential, confined to the cation sphere, and acting only upon the $d$
radial wavefunctions. The unoccupied n$d$ states are then pushed higher
in energy. The calculated effective charges, for the whole series of
fake materials, are reported in Table~\ref{table2}. As expected, our
modification of the potential has no effect on the calculated $Z^*$ of
MgO; instead the effects are quite relevant for the other three cases,
where the values of $Z^*$ are much closer to the nominal ionic charges
than in the real materials. Almost all partial contributions change
conspicuously, not only those corresponding to states closest to the
fundamental gap, since the fake repulsive potential affects the lowest
bands as well. However, the partial variations corresponding to low
lying states are unimportant, since they are due to intermixing of
occupied states amongst themselves. In this respect the situation is
the same for the fake crystals as for the real ones: the sum of the O
$2s$ and M n$p$ contributions adds up to an almost nominal value. The
only important variation is in the contribution of the O $2p$
electronic states, which can no longer interact with the lowest
conduction states, thus resulting in a near--nominal value of the
dynamical charges.

Finally, we note an heuristic trend in the partial contributions given
in Tables~\ref{table1} and \ref{table2}, as well as those available in
the literature for the dynamical charges.\cite{Ghosez95,rap88} In the
case of an anion displacement, when $p$ states interact with $s$ ($d$)
states, the deviations from the nominal value of the partial
contributions to $Z^*$ are positive (negative) for $p$ states, and
negative (positive) for $s$ ($d$) states. In the case of a cation
displacement, the above effects are opposite. This rule controls also
the total value $Z^*$ of the three heaviest oxides, where the only
relevant interaction is the one between the highest valence states
(predominantly O $2p$--like) and $d$--like conduction states.

We thank N. Binggeli for her interest in this work. We are also
grateful to D. Singh for helpful discussions concerning the
local--orbital method. This work was supported by the Swiss National
Science Foundation grant 20--39.528.93. H.K acknowledges support by
NSF grant DMR-9404954, and R.R acknowledges partial support by ONR
grant N00014-96-1-0689. Calculations have been performed on the
computers of EPF--Lausanne and ETH--Z\"urich and the Cornell Theory
Center.

\begin{figure}
\caption[1]{Energy bands of MgO calculated (a) at the experimental
value a=7.97 a.u. of the lattice parameter and (b) under negative
pressure, corresponding to a=9.09 a.u. as in CaO.}
\label{fig1}
\end{figure}
\begin{figure}
\caption[2]{Energy bands of (a) CaO, (b) SrO, and (c) BaO calculated at
the experimental value of the lattice constant.}
\label{fig2}
\end{figure}

\begin{table}
\caption[t1]{Decomposition of the Born dynamical charge $Z^*_{\rm O}$
of oxygen for the series of alkaline--earth oxides into contributions
from different groups of bands and from the ionic cores. For SrO, the
contributions from O $2s$ and Sr $4p$ cannot be separated; their sum is
$-$2.06. The nominal values, in a completely ionic picture, are given
in brackets.}
\begin{tabular}{ldddd}
                & MgO & CaO & SrO & BaO \\ \tableline
M n$s$ \hspace{0.2cm}[0.0]    & $-$0.02 & $-$0.01 & $-$0.01 & +0.00 \\
O 2$s$ \hspace{0.2cm}[$-$2.0] & $-$2.21 & $-$2.68 &         & $-$2.72 \\
M n$p$ \hspace{0.2cm}[0.0]    & +0.01   & +0.59   &         & +0.74 \\
O 2$p$ \hspace{0.2cm}[$-$6.0] & $-$5.85 & $-$6.35 & $-$6.46 & $-$6.82 \\
Core \hspace{0.2cm}[+6.0]     & +6.00   & +6.00   & +6.00   & +6.00 \\ 
\tableline
Total \hspace{0.2cm}[$-$2.0]  & $-$2.07 & $-$2.45 & $-$2.53 & $-$2.80 \\ 
\tableline
$Z^*_{\rm O}$ Expt.\tablenotemark[1] 
                     & \multicolumn{1}{c}{1.96/2.02}
                     & \multicolumn{1}{c}{2.26/2.30}
                     & \multicolumn{1}{c}{2.34/2.47}
                     & \multicolumn{1}{c}{2.69/2.97} \\
\end{tabular}
\label{table1}
\tablenotetext[1]{From Ref.~\onlinecite{landolt82}}.
\end{table}

\begin{table}
\caption[t2]{Same as in Table~\ref{table1}, but with a fake repulsive
potential acting, within the cation sphere only, on the $d$--like
component of the electronic states. For SrO, the sum of the O $2s$ and
Sr $4p$ contributions is $-$2.02}
\begin{tabular}{ldddd}
            & MgO & CaO & SrO & BaO \\ \tableline
M n$s$      & $-$0.03 & $-$0.04 & $-$0.04 & $-$0.05 \\
O 2$s$      & $-$2.22 & $-$2.37 &         & $-$3.03 \\
M n$p$      & +0.02   & +0.32   &         & +1.07 \\
O 2$p$      & $-$5.83 & $-$5.91 & $-$6.03 & $-$6.25 \\
Core        & +6.00   & +6.00   & +6.00   & +6.00 \\ 
\hline
Total       & $-$2.06 & $-$2.00 & $-$2.09 & $-$2.26 \\
\end{tabular}
\label{table2}
\end{table}
\end{document}